\documentclass[10pt,letterpaper]{article}
\usepackage[top=0.85in,left=2.75in,footskip=0.75in]{geometry}

\usepackage{amsmath,amssymb}

\usepackage{changepage}

\usepackage[utf8x]{inputenc}

\usepackage{textcomp,marvosym}

\usepackage{cite}

\usepackage{nameref,hyperref}

\usepackage{microtype}
\DisableLigatures[f]{encoding = *, family = * }

\usepackage[table]{xcolor}

\usepackage{array}

\usepackage{caption}
\usepackage{subcaption}
\usepackage{amsthm}
\newtheorem{theorem}{Theorem}
\newtheorem{lemma}{Lemma}
\newtheorem{corollary}{Corollary}
\usepackage{xcolor}

\newcolumntype{+}{!{\vrule width 2pt}}

\newlength\savedwidth

\raggedright
\usepackage[aboveskip=1pt,labelfont=bf,labelsep=period,justification=raggedright,singlelinecheck=off]{caption}

\makeatletter
\renewcommand{\@biblabel}[1]{\quad#1.}
\makeatother

\usepackage{lastpage,fancyhdr,graphicx}
\usepackage{epstopdf}
\fancyheadoffset[L]{2.25in}
\fancyfootoffset[L]{2.25in}
\begin{document}
\vspace*{0.2in}

\begin{flushleft}
{\Large
\textbf\newline{Axioms for Clustering Simple Unweighted Graphs: no impossibility result} %
}
\newline
\\
James Willson\textsuperscript{1},
Tandy Warnow\textsuperscript{1*}
\\
\bigskip
\textbf{1} Department of Computer Science, University of Illinois at Urbana-Champaign, Urbana, IL, USA
\\
\bigskip

*warnow@illinois.edu

\end{flushleft}
\section*{Abstract}
In  2002, Kleinberg proposed three axioms for distance-based clustering,    and proved that it was impossible for a clustering method to satisfy all three. While there has been much subsequent work examining and modifying these axioms for distance-based clustering,  little work has been done to explore axioms relevant to the graph partitioning problem when the graph is unweighted and given without a distance matrix. Here, we propose and explore axioms for graph partitioning for this case, including  modifications of Kleinberg's axioms and three others: two axioms relevant to the ``Resolution Limit'' and one addressing well-connectedness.  We prove that clustering under the Constant Potts Model satisfies all the axioms, while Modularity clustering and iterative $k$-core both fail many axioms we pose.
These theoretical properties of the clustering methods are relevant both for theoretical investigation as well as to practitioners considering which methods to use for their domain science studies.

\section*{Author summary}
\textcolor{black}{
In 2002, Kleinberg proposed three axioms for distance-based clustering and proved that it was not possible for any clustering method to simultaneously satisfy all three axioms. 
 Here, we examine these axioms in the context where the input network is given without any pairwise distance matrix and is instead a simple unweighted graph.  For this case we propose corresponding axioms, and we we include three  additional axioms, two related to the resolution limit and the other related to well-connectedness. We establish that some methods, such as optimizing under the Constant Potts Model,  satisfy all the axioms we pose, but that others (notably clustering under the Modularity optimization problem) fail to satisfy some of these axioms.
 This study sheds light on limitations of existing clustering methods.} 


\section*{Introduction}
Graph clustering, also known as community detection or graph partitioning, is the problem of taking a graph $G=(V,E)$ as input and returning a partition of the vertex set into disjoint subsets, referred to equally as clusters or communities. 
In some contexts, the graph is given as a distance matrix $D$ so that $D[i,j]$ is the distance between vertices $i$ and $j$.

In 2002, Kleinberg \cite{kleinberg2002impossibility} defined three axioms (Richness, Consistency, and Scale-Invariance) for clustering based on distances, and  proved that it was impossible for any clustering method to satisfy all three axioms.   
The Refinement-Consistency axiom is a relaxation of Consistency, but Kleinberg \cite{kleinberg2002impossibility} also proved an impossibility result with this substitution. 

The apparent impossibility of distance-based clustering to satisfy all stated desirable axioms drove research in several directions.
For example, \cite{ackerman2012towards} addresses the Consistency axiom, pointing out cases where it might not be   desirable. 
Furthermore, there has been work in sidestepping axioms by defining the number of clusters in advance  \cite{zadeh2012uniqueness,cohen2018clustering}. For example, \cite{zadeh2012uniqueness} does this by replacing Richness with $k$-Richness, which is a version of Richness restricted only to consider clusterings with $k$ clusters, and \cite{cohen2018clustering} argue that Consistency should not hold if the ``correct'' number of clusters changes.  
Additional work has also been done applying the principles of Kleinberg's  distance-based axioms to quality measures instead of directly to the clustering function. 
For example, \cite{ben2008measures} formulates such a set of axioms and  shows that these new axioms do not lead to an impossibility result.

Here we consider axiomatic properties of clustering when the input is an \textcolor{black}{unweighted simple graph  (i.e., neither the edges nor the vertices are weighted) and where the graph is given without  a distance matrix.  We also assume the number of clusters is not known in advance.
The motivation for considering these simple unweighted graphs is that many real-world graphs are of this form (e.g., citation graphs).}  In addition, while it is certainly {\em possible} to define a pairwise distance matrix relating the vertices (e.g., the length of the shortest path between each pair of vertices), such approaches lose information about the input graph (see discussion in \cite{schaeffer2007graph,fortunato2010community}). Finally, graph clustering 
when the input does not include a distance matrix is very common (e.g., see the DIMACS report \cite{bader2013graph}).

Very little has been done to discuss axiomatic approaches for graph clustering when the input is a graph without any distance matrix.
However, three studies \cite{schaeffer2007graph,fortunato2010community,van2014axioms} provide overviews of the literature related to axiomatic properties of clustering methods, with Kleinberg's axioms reformulated for the distanceless case. 
Of these, \cite{van2014axioms} provides theoretical advances in axiomatic properties of clustering methods when the input graph has non-negative edge-weights, and established that Modularity-optimization satisfies Richness.

\textcolor{black}{Another property that has been discussed in the literature is the ``resolution limit" \cite{fortunato2007resolution}, which roughly speaking indicates that a clustering method has a lower bound on the size of the clusters it can find.
This resolution limit was established for Modularity optimization in \cite{fortunato2007resolution}, using a ring-of-cliques as an example of how Modularity can fail to find the obvious communities (i.e., the cliques) as the number of cliques grows but not their size.
This observation led to the development of other methods, including an approach to clustering based on optimizing under the Constant Potts Model \cite{traag2011narrow}, for which the failure on the ring-of-cliques example does not hold. 
}

\textcolor{black}{
We expand on the prior work by formulating seven axioms suitable for   clustering methods that operate on unweighted simple networks.}
Four of these axioms  are reformulations of  Kleinberg's original  Richness and Consistency axioms, following on \cite{schaeffer2007graph} and \cite{fortunato2010community}  for the distanceless case.
\textcolor{black}{
The final three axioms  include one that addresses how well-connected the clusters are (i.e., considering the size of the minimum edge cut of each cluster) and two others that are related to the resolution limit, one of which was formulated in \cite{traag2011narrow}.} 
We find that CPM-optimization satisfies all the axioms we pose, but all other clustering methods we study, including Modularity optimization,  
fail to satisfy most of the axioms we pose.

Our study provides new evidence that CPM-based optimization has superior theoretical properties compared to Modularity-optimization. 
It also sheds light on the tricky question of which methods suffer from the ``resolution limit'', as the original formulation in \cite{fortunato2007resolution} and the response from \cite{traag2011narrow} do not fully overlap. 
In addition to proposing new research questions for theoreticians, the insights from this study provide useful insight for domain scientists in selecting methods for use in their empirical work.

\section*{Background}

\subsection*{Clustering methods}
We discuss theoretical properties of Modularity,  CPM (constant Potts model), and IKC (iterative $k$-core) clustering. 
\textcolor{black}{We also consider two ``toy" clustering methods: 
\begin{itemize}
    \item Components-are-Clusters: the clustering method that returns the connected components of the network as the clusters 
    \item Nodes-are-Clusters: the clustering method that returns every node as a singleton cluster
\end{itemize}
}

\subsubsection*{Modularity}
Modularity, introduced in \cite{newman2004finding}, is an optimization problem that we now define. 
Given a clustering $\mathcal{C}$ of $N$, 
we define the Modularity score of $\mathcal{C}$ as follows.
$E$ denotes the set of edges in the network $N$, 
$e_c$ is the number of edges internal to cluster $c$, and $d_c$ is the sum of the degrees of nodes  found in cluster $c$ (noting that the degree of a node $v$ in a cluster $c$ is the total number of neighbors of $v$, whether or not in the cluster).
The Modularity score of $\mathcal{C}$ is 
\begin{equation}
    \mathcal{H} = \sum_{c \in \mathcal{C}}\left[\frac{e_c}{|E|} - \left(\frac{d_c}{2|E|}\right)^2\right].
    \label{eq:mod}
\end{equation}

The Modularity optimization problem, which takes as input a network and  seeks a clustering with the largest modularity score,  was proven NP-hard in \cite{brandes2007modularity}. 
We make a minor modification to the Modularity optimization problem by requiring that the  clusters be connected. 

\subsubsection*{The Constant Potts Model (CPM) clustering problem} 
Optimizing under the
Constant Potts Model (CPM) \cite{traag2011narrow} was developed as a way of addressing two  weakness in Modularity optimization that it is subject to the resolution limit \cite{fortunato2007resolution}.
The CPM optimization criterion takes a parameter $\gamma$ (the resolution value).
Letting  $e_c$ denote the number of edges and $n_c$  the number of nodes in cluster $c$, the CPM score of    $\mathcal{C}$ is  
\begin{equation}
    \mathcal{H} = \sum_{c \in \mathcal{C}}\left[e_c - \gamma\binom{n_c}{2}\right].
    \label{eq:cpm}
\end{equation}

Note that the optimization problem depends on the resolution parameter $\gamma$; in this study, we will constrain $\gamma$ to be in the open interval $(0,1)$.
When not clear by context, we refer to the  usage of CPM with a fixed value for parameter $\gamma$ as CPM($\gamma$).

\subsubsection*{IKC and IKC(no-mod)}
The iterative $k$-core \cite{wedell2022center} algorithm (also known as IKC) is a deterministic clustering algorithm based on finding $k$-cores, which are maximal connected subgraphs where every vertex is adjacent to at least $k$ other vertices in the subgraph. 
A $k$-core can be found by iteratively pruning all nodes with degree smaller than $k$ from the graph until no more remain. 
IKC operates by determining the largest $k$ for which a $k$-core exists, removes that $k$-core, and then recurses. 
IKC takes a parameter $k_0$ and only returns those clusters that satisfy two properties: the minimum degree within the cluster is at least $k_0$ and every non-singleton cluster has positive Modularity score.
In this study, we consider two versions of IKC: both have $k_0=0$ and one drops the requirement of positive Modularity for each non-singleton cluster. 
We refer to the version that drops Modularity as IKC(no-mod) and the other as simply IKC.

\subsection*{Kleinberg's axioms} 
In distance-based clustering,   a clustering function $f$ takes set $S$ with $n$ elements and an $n \times n$ distance matrix $d$ and returns $\Gamma$, which is a partition of $S$. 
With this notation, \cite{kleinberg2002impossibility} proposed the following three axioms:
\begin{itemize}
\item 
 \textbf{Scale Invariance:} Given some constant $\alpha > 0$, $f(d) = f(\alpha \cdot d)$.
 In other words, if all the distance between points in the data are multiplied by a constant amount this should not affect the output of the clustering method. 
 \item 
 \textbf{Richness:} 
 \textcolor{black}{The clustering function $f$ satisfies, for all networks and clusterings $\Gamma$, that there is some distance matrix $d$ on the network such that $f(d) = \Gamma$.
In other words, there should not be any clustering that is impossible to obtain. } 
 \item 
\textbf{Consistency:} Given two distance functions $d$ and $d'$, $f(d) = f(d')$ if $d'$ transforms $d$ in the following way: If $i$ and $j$ are from the same cluster then $d'(i,j) \leq d(i,j)$; otherwise, if they are from different clusters $d'(i,j) \geq d(i,j)$. This stands to reason, as if the clusters are made tighter, or if the clusters are made more distinct from one another (by being moved further away from each other), then it seems as if these changes should reinforce the existing clustering.
\item  
 \textbf{Refinement-Consistency:} This is the same as Consistency except for the following change: instead of requiring that $f(d) = f(d')$, it is sufficient that every cluster in $f(d')$ is a subset of a cluster in $f(d)$.
Kleinberg's study showed that   his impossibility result held even with this relaxation.  
\end{itemize}

\subsection*{The Resolution Limit}

\textcolor{black}{
As shown by Fortunato and Barth\'elemy in 
\cite{fortunato2007resolution}, Modularity optimization can fail to return what are obvious true communities if they are too small. Specifically, Fortunato and Barth\'elemy described an infinite family of networks formed of rings of cliques, each clique connected to  each of its two neighbors by a single edge, where the cliques are a constant size but the number of the cliques increases.  Fortunato and Barth\'elemy prove that if the number the  cliques is large enough, then Modularity will stop returning the cliques as communities and will instead return sets of cliques as communities. They   described this by saying that Modularity suffers from the resolution limit.}

Traag {\em et al.} \cite{traag2011narrow} proposed the following definition of what it means for an optimization problem (or  method that solves the optimization problem exactly) to be ``resolution-limit free":
\emph{Let $\mathcal{C} = \{C_1, C_2, \ldots, C_q\}$  be a $\mathcal{H}$-optimal partition of a graph $G$. Then the objective function $\mathcal{H}$ is called \textit{resolution-limit-free} if for each subgraph $H$ induced by $\mathcal{D} \subset \mathcal{C}$, the partition $\mathcal{D}$ is also $\mathcal{H}$-optimal.} 
\cite{traag2011narrow} prove that, according to this definition,  optimizing under the Constant Potts Model (CPM) is resolution-limit-free but  optimizing under the Modularity criterion is not resolution-limit-free.

Of concern to us, in this study, is that this definition of resolution-limit-free does not  address in full the issue raised by \cite{fortunato2010community}.
For example, \textcolor{black}{a method that returns each component in the network as a cluster satisfies the definition of ``resolution-limit-free'' as provided by \cite{traag2011narrow} but fails to return the cliques inside the ring-of-cliques component as communities and will instead return the entire component.}

\subsection*{Well-connectedness}
A natural expectation of a community (i.e., cluster) is that it should be both dense (i.e., have more edges inside the cluster than would be expected by chance) and well-connected (i.e., not have a small edge cut). 
However, definitions for ``well-connected'' vary by study.  
For example,  \cite{traag2019louvain} established a lower bound on the cut size for a CPM-optimal clustering as a function of the resolution parameter $\gamma$, so that if an edge cut splits a cluster into two sets $A$ and $B$ then the edge cut has size at least $\gamma \times |A| \times |B|$, and used this as the definition for ``well-connected'' clusters.
\cite{park2023identifying-conf} showed empirically that many clustering methods, including CPM-clusterings produced using the Leiden \cite{traag2019louvain} software, often produced
clusters with small edge cuts, and even produced clusters that were trees.  
Based on this observation, 
\cite{park2023identifying-conf} proposed instead that a cluster be considered well-connected if the size of a min cut in a cluster with $n$ nodes is greater than $f(n)$, where $f(n)$ is a non-decreasing function provided by the user that increases to infinity.

\section*{Our distanceless axioms}
In the distanceless context, our input is a simple unweighted undirected graph 
$N = (V, E)$, where $V$ is the vertex set and $E$ is the edge set.
We propose seven axioms, where the first four are obtained by modifying Kleinberg's axioms for the distanceless context, one is designed to address well-connectedness, and a final two relate to the resolution limit (one introduced earlier in \cite{traag2011narrow}).

\begin{itemize}
    \item 
 \textbf{Richness:} A clustering method $M$ satisfies richness if, for any clustering $\Gamma$ of a set $V$,    there exists an edge set $E$ so that  $M(N)=\Gamma$ when  $N= (V,E)$. \textcolor{black}{Note that we allow for the trivial clusterings, i.e., when all the nodes are in the same cluster, or when they are each in separate clusters.} \\ 

        \item  \textbf{Standard Consistency:}  
        A clustering method $M$ satisfies standard consistency if, for every graph $N=(V,E)$  and output clustering $M(N)$, when $E'$ differs from $E$ by the removal of edges between clusters in $M(N)$ or the addition of edges within clusters in $M(N)$, then $M(N')=M(N)$ where $N'=(V,E')$.\\
        
        \item  \textbf{Refinement Consistency:} 
        This is a relaxation of Standard Consistency where adding internal edges to a cluster is allowed to split the cluster apart but no other changes are allowed.\\
        
\item \textbf{Inter-edge Consistency:} This is a relaxation of Standard Consistency, where the clustering must remain unchanged when edges between clusters are removed.\\

\item \textbf{Connectivity:} 
\textcolor{black}{
We  extend \cite{park2023identifying-conf} to define this axiom. 
We say that a 
cluster is  well-connected if the size of the minimum edge cut exceeds $f(n)$, where $n$ is the number of nodes in the cluster and $f$ is an arbitrary non-negative non-decreasing function that approaches infinity.    
We say that a clustering method $M$ satisfies connectivity if and only if for some function
$f:\mathbb{R^+} \rightarrow \mathbb{R^+}$ that is non-negative, non-decreasing, and that approaches infinity, 
for all networks $N$ and all non-singleton clusters $c$ in the  clustering produced by $M$, $c$ is well-connected.  }\\

\item 
   \textbf{Pair-of-Cliques: } 
   This axiom is a small start towards a more thorough evaluation of robustness to the resolution-limit, since the characterization in \cite{traag2011narrow} does not adequately address the concerns raised in \cite{fortunato2007resolution}.
Recall that \cite{fortunato2007resolution} presented the resolution limit problem with an example of a network containing a ring of $n$-cliques, and established that as the number of cliques increased Modularity optimization would fail to return the cliques as communities, returning instead clusters containing two or more of these cliques.
Since a ring of cliques is not the only condition where methods can fail to detect small or meso-scale communities, we
consider a simple case   where one component in the network contains a pair of $n$-cliques, connected by an edge, and we refer to this as a Pair-of-Cliques component.
We say a graph partitioning method  satisfies the Pair-of-Cliques axiom if \textcolor{black}{there is a constant $n_0$ such that if the network $N$ has a Pair-of-Cliques component of size at least $n_0$  then  the clustering method would return $A$ and $B$ as separate clusters.}
\\

\item 
\textbf{Fixed-Point: }
\textcolor{black}{
We consider the property proposed in \cite{traag2011narrow}, whereby a method is said to be  ``resolution-limit-free" if iteratively applying the clustering method will not change the clustering.  We refer to this as satisfying the Fixed-Point axiom. }

\end{itemize}
\begin{table}[!ht]
\begin{adjustwidth}{-2.25in}{0in} %
\small
\centering
\caption{
{\bf Overview of Theoretical Results.}}
\begin{tabular}{l|ccccccc}
    \hline
    Method & Richness & Std Consist. & Ref Consist. & Inter-Edge Consist. & Connect & Pair-of-Cliques & Fixed Point\\
    \hline 
  Components & \checkmark \cite{van2014axioms}& \checkmark & \checkmark & \checkmark &-  & - & \checkmark\\
  Nodes & - & \checkmark &\checkmark & \checkmark & \checkmark & - & \checkmark \\
    CPM($\gamma$)  & \checkmark %
    & \checkmark & \checkmark & \checkmark & \checkmark  & \checkmark & \checkmark \cite{traag2011narrow} \\
     Modularity & \checkmark \cite{van2014axioms} & - & - & - & -  & - \cite{fortunato2007resolution} & - ~\cite{traag2011narrow}\\
    IKC & - &  - & - & - &  - & - & -\\
    IKC(no-mod) &  \checkmark & - & - & \checkmark &  - & -& \checkmark\\
    \hline
\end{tabular}
\begin{flushleft} For each clustering method, we show which axioms are satisfied, with a reference to the paper where the result was first established if not in this study. A \checkmark indicates that the method follows the axiom and ``-'' indicates the method fails to follow the axiom. In CPM($\gamma$), we assume $\gamma$  (the resolution  parameter) is arbitrary but fixed. IKC(no-mod) is the variant of IKC where the requirement that non-singleton clusters have positive Modularity is dropped.
The first four axioms are modifications of Kleinberg's axioms for the distanceless case, the next two are new axioms we introduce, and the final one is our name for the property referred to as ``resolution-limit-free" from \cite{traag2011narrow}.
\end{flushleft}
\label{tab:results_summary}
\end{adjustwidth}
\end{table}

\section*{Results}

In some cases we provide sketches of proofs, leaving full proofs to Appendix.
We begin with a lemma.
\textcolor{black}{
\begin{lemma}
If $M$ is a clustering method that satisfies Connectivity, then for some $n_0 \geq 1$, no clusters returned by $M$ of size at least $n_0$ have cut edges.
\label{lemma:connectivity-no-cutedge}
\end{lemma}
\begin{proof} 
Suppose $M$ satisfies Connectivity. 
By definition, there is some function $f$ that is non-decreasing and satisfies $f(x) \rightarrow \infty$ as $x \rightarrow \infty$, such that for all networks $N$ and all clusters $C$ returned by $M$ on $N$, the min cut size of $C$ is strictly greater than $f(n)$ where $n$ is the size of $C$.
Since $f(x)$ is non-decreasing and converges to infinity, there is some $n_0$ so that for all $n \geq n_0, f(n) \geq 1$. 
Hence, for all networks $N$ and clusters of size at least $n_0$ returned by $M$ on $N$, the mincut size for the cluster will be strictly greater than $1$, and so no found cluster of size at least $n_0$ can have a cut edge.
\end{proof}
    }

 \subsection*{Theory for Components-are-Clusters}
 \textcolor{black}{
Recall that the Components-are-Clusters method returns every connected component as a cluster.
\label{thm:components-theory}
\begin{theorem}
Components-are-Clusters satisfies the Richness, Standard Consistency, and Fixed Point axioms, but fails Connectivity and the Pair-of-Cliques axioms. 
\end{theorem}
}
\begin{proof}
\textcolor{black}{
    First we establish  Richness. Suppose we are given clustering $\Gamma$ of a set $V$ of nodes. For every cluster in $\Gamma$, we make all the nodes in the cluster pairwise-adjacent, i.e., each cluster now becomes a clique. 
    No other edges are added, so that every cluster is a connected component in the network. 
    Components-are-Clusters will return each connected component as a cluster,  and thus satisfies Richness.} 
    
    For Standard Consistency, note that adding edges between nodes in a connected component can never connect two disconnected components, nor can it split a component. The same is true for removing edges between two connected components. 
    Thus, Components-are-Clusters satisfies Standard Consistency.

    For Connectivity, the proof is by contradiction. 
 If Components-are-Clusters satisfied Connectivity, then by Lemma \ref{lemma:connectivity-no-cutedge}, there is some $n_0$ such that no cluster of size at least $n_0$ returned by Components-are-Clusters can have a cut edge.  Now consider  a network $N$ that has a component of size $n_0$ that is a tree.
 Components-are-Clusters would return that component, thus failing Connectivity.

    For the Pair-of-Cliques axiom, any component consisting of  a pair of cliques connected by a single edge would be returned by Components-are-Clusters.  Since we can make such a component arbitrarily large,  Components-are-Clusters fails the Pair-of-Cliques axiom. 

\textcolor{black}{
    Finally, consider the Fixed Point axiom. 
    Clearly, applying Components-are-Clusters to any network twice will return exactly the same clustering, and so Components-are-Clusters satisfies the Fixed Point axiom.}
\end{proof} 
\subsection*{Theory for Nodes-are-Clusters}
\textcolor{black}{
Recall that the Nodes-are-Clusters method returns every node as a singleton cluster.
\label{thm:nodes-theory}
\begin{theorem}
Nodes-are-Clusters fails Richness and Pair-of-Cliques and satisfies Connectivity, Standard and Refinement Consistency, and the Fixed Point axioms.
\end{theorem}
}
\begin{proof}
Nodes are clusters will return $n$ clusters given any network on $n$ nodes, and so fails the Richness axiom.
Similarly, it fails Pair-of-Cliques, as it cannot return any clique of size greater than $1$ as a cluster.
The connectivity axiom is  satisfied, since letting the axiom is only applied to non-singleton clusters.  
Standard consistency follows, since adding or deleting edges from a network does not change the clustering.
\textcolor{black}{Similarly, Nodes-are-Clusters trivially satisfies the Fixed Point axiom,  since applying Nodes-are-Clusters to any network twice will return exactly the same clustering.}
\end{proof} 
\subsection*{Theory for CPM}

\begin{theorem}
For all values $\gamma >0$,
CPM($\gamma$) follows all axioms. 
      \label{thm:cpm-theory}
\end{theorem}

That CPM($\gamma$) satisfies the Fixed Point axiom was established in \cite{traag2011narrow}.
We now provide proofs that CPM($\gamma$) follows the remaining axioms, assuming in each case that $0< \gamma < 1$ is fixed but arbitrary.

\begin{lemma}

    CPM($\gamma$) is Rich.
    \label{cpm-rich}
\end{lemma}
\begin{proof}
Let $V$ be a set of nodes and $\Gamma$ a partition of $V$.
For each set in the partition, form a clique.
For any $\gamma >0$, the clustering that puts every clique into a cluster attains the largest possible score, and all other clusterings have lower scores.
Thus, CPM($\gamma$) satisfies the Richness axiom.
\end{proof}

\begin{lemma}
    CPM($\gamma$) follows Inter-Edge Consistency.
    \label{lemma:cpm-interedge}
\end{lemma}

\begin{proof}

Let $\gamma>0$ be fixed, and let $G=(V,E)$ be a network.
    Let $\Gamma$ be a  clustering $\{c_1, c_2, \cdots, c_m \}$   of $G$ that is CPM-optimal.
    Let $E'$ be a subset of $E$ produced by removing some edges whose endpoints are in different clusters in $\Gamma$. 
    We let $CPM(c, E)$ denote the CPM score for cluster $c$ given edge set $E$ 
    \textcolor{black}{and  $e'_c$ denote the number of edges from $E'$ in $c$.}
    From Equation \ref{eq:cpm}, we see that
    \begin{align*}
        \forall c_i \in \Gamma,\; &CPM(c_i, E) = CPM(c_i, E').\\
        \intertext{Additionally,}\\
        \forall c \in \mathcal{P}(V),\; &CPM(c, E) \geq CPM(c, E'),\\
        \intertext{where $\mathcal{P}(V)$ is the power-set of $V$, as}\\
        \forall c \in \mathcal{P}(V),\; &e_c \geq e'_c\\
    \end{align*}
Therefore, $\Gamma$ remains optimal.
\end{proof}

\begin{lemma}
    CPM($\gamma$) follows Standard (and therefore Refinement) Consistency.
    \label{lemma:cpm-standard-consistency}
\end{lemma}

\begin{proof}

Let $\gamma>0$ be fixed, and let $G=(V,E)$ be a network.
    Consider an optimal clustering $\Gamma$ and imagine adding a single edge into one of the clusters. 
    The score of $\Gamma$ will go up by 1, since the edge was added to a cluster within $\Gamma$. 
    As per Equation \ref{eq:cpm}, the most that the CPM score of any other clustering can increase by is exactly 1; hence $\Gamma$ remains optimal after adding that edge. Therefore, inductively, a clustering that is optimal for a network given edge set $E$ remains optimal if we add edges within the clusters.
    We also note that removing edges does not need to be considered, as CPM was shown to satisfy inter-edge consistency in Lemma \ref{lemma:cpm-interedge}.   
\end{proof}

\begin{lemma}
    CPM($\gamma$) is Connective.
    \label{lemma:cpm-gamma-connective}
\end{lemma}

\begin{proof}
A proof of this theorem also follows from Equation D1 in the Supplementary Information in \cite{traag2019louvain}; here we provide a simple proof.

Given $\gamma>0$, we let function $f_{\gamma}$ be defined
by $f_{\gamma}(n)= \lceil \gamma (n-1) \rceil$.
Note that $f$ maps positive integers to integers, is non-decreasing, and grows unboundedly (i.e., $f_{\gamma}(n) \rightarrow \infty$ as $n \rightarrow \infty$). 
We will show that for every $\gamma$, every network $N$,  and every $CPM(\gamma)$-optimal clustering of $N$, the minimum edge cut of any cluster $c$ in the clustering is at least size $f_{\gamma}(n_c)$, where $n$ is the number of nodes in the cluster $c$.
Therefore, this will establish that CPM($\gamma$) is Connective.

Suppose $C$ is a cluster with $n$ nodes in a CPM-optimal clustering of a network $N$ for some fixed $\gamma$.  We   consider an edge cut  $E_0$ for $C$. 
Since $C$ is a cluster in a CPM-optimal clustering, dividing $C$ into two clusters  cannot improve the CPM-score.
Hence, whatever  division of $C$ into two sets is produced by deleting $E_0$, the best that can happen is that the CPM-score is not reduced.

Let  $n'$ denote the number of nodes on one side of the edge cut, $A$ denote the number of edges connecting those nodes, and $B$ denote the number of edges connecting the nodes on the other side of the edge cut.
The CPM score of cluster $C$ is  $ A + B + |E_0| - \gamma \binom{n}{2}$. Therefore,  we obtain
    \begin{align*}
        A + B + |E_0| - \gamma \binom{n}{2} &\geq A - \gamma \binom{n'}{2} + B - \gamma \binom{n-n'}{2},\\
        \end{align*}
       because the score of the cluster $C$ is at least  the sum of the scores of the subclusters produced by deleting $E_0$.
       Therefore,
        \begin{align*}
        |E_0| &\geq \gamma \binom{n}{2} - \gamma \left[ \binom{n'}{2} + \binom{n-n'}{2} \right].
    \end{align*}
    We then note
    \begin{align*}
        |E_0|  &\geq \gamma \binom{n}{2} - \gamma \max_{n'} \left[ \binom{n'}{2} + \binom{n-n'}{2} \right]\\
        &= \gamma \left[ \binom{n}{2}  - \binom{n-1}{2} - \binom{1}{2}\right]\\
        &= \gamma \left[ \binom{n}{2}  - \binom{n-1}{2}\right]\\ 
        & = \gamma (n-1).
    \end{align*}
    Hence, $|E_0| \geq f(n,\gamma)$ for any
    edge cut $E_0$ separating a CPM$(\gamma)$-optimal cluster $C$ with $n$ nodes.
\end{proof}

\textcolor{black}{
In the Appendix, we examine the guarantee that CPM($\gamma$) is connective in greater detail.  
Specifically, Lemma \ref{lemma:appendix-cpm-return-component-small-gamma} establishes  that the connectivity guarantee provided for CPM($\gamma$) depends on $\gamma$, and that small values for $\gamma$ allow for large clusters with cut edges being returned.}

\begin{lemma}
    CPM($\gamma$) satisfies the Pairs-of-Cliques axiom.
\end{lemma}
\begin{proof}
To show that CPM($\gamma$) satisfies the Pairs-of-Cliques axiom, we must show that for a fixed $\gamma>0$, there is value for $n$ where all cliques with $n$ vertices or more will be clustered in separate clusters if connected by a single edge. 
Since CPM($\gamma$) is connective,   we can pick $N$ large enough so that the mincut size for any  cluster of size at least $N$ is at least two. 
Hence,  if $C_1$ is a component in $N$ that has two $n$-cliques connected by an edge and $2n \geq N$, then no clustering of $C_1$ in a CPM-optimal clustering can have a cut edge. 
Therefore, each of the clusters of $C_1$ in an optimal CPM clustering must be subsets of $A$ or $B$.
It is easy to see that the CPM score is maximized by returning $A$ and $B$ as clusters, and so CPM($\gamma$) follows the Pair-of-Cliques axiom.
\end{proof}

    \subsection*{Theory for Modularity}

\begin{theorem}
    Modularity follows Richness, but violates all the other axioms, i.e., the Standard and Refinement Consistency, Inter-edge Consistency, Connectivity, Pair-of-Cliques, and Fixed Point axioms.
    \label{theorem:modularity}
\end{theorem}
\begin{proof} 

   Modularity was shown to satisfy Richness in Theorem 1 (with the proof  in Appendix A) of \cite{van2014axioms}, and 
   was shown to fail the Fixed Point axiom in \cite{traag2011narrow}. 

      We now sketch the proof that Modularity violates Refinement Consistency and hence Standard Consistency (see Appendix Lemma \ref{lemma:appendix-modularity-standard-consistency} for full details). 
      In Appendix Lemma \ref{lemma:appendix-modularity-standard-consistency}, we consider a network $N$ that has a component  $G_1$ that is a pair-of-cliques (i.e., it has two node-disjoint $n$-cliques (with $n \geq 5$) $A$ and $B$ that are connected by an edge).
    Appendix Lemma \ref{lemma:appendix-no-split-cluster} establishes that a Modularity-optimal clustering of $N$ will either return   $G_1$  as a cluster or will return the two $n$-cliques $A$ and $B$ as clusters.
     In Appendix Lemma \ref{lemma:appendix-modularity-standard-consistency}, we then consider a network $N$ with  $G_1$ as one component and with a second component $G_0$ that  is a $p$-star (i.e., the graph with a single node adjacent to $p$ other nodes, and no other edges).  
     Appendix Lemma \ref{lemma:appendix-modularity-interedge} shows that for $n \geq 5$ and $p$ large enough,   the Modularity-optimal clustering of the network will produce $A$ and $B$ as two clusters, and that  when $G_0$ is  a $(p+1)$-clique  then a Modularity-optimal clustering will return $G_1$ as a cluster. Thus, adding edges within a cluster can change the clustering.
      This shows that Modularity violates Refinement Consistency, which in turn establishes that it violates Standard Consistency.
      Note that this argument also establishes that Modularity violates the Pair of Cliques axiom.

      The proof that Modularity violates Inter-edge Consistency is provided in Appendix Lemma \ref{lemma:appendix-modularity-interedge} and uses a similar argument to Lemma \ref{lemma:appendix-modularity-standard-consistency}. 
      \textcolor{black}{
      We construct a graph with two components, where the first component is a pair-of-cliques component. In Lemma \ref{lemma:appendix-modularity-interedge}, we show that this network    has the following two properties: The optimal modality clustering of the network containing both components returns the pair-of-cliques component as a single cluster and splits the other component into multiple clusters, and  if any edge is removed from the second component, then the first property is no longer satisfied. Hence, Modularity violates Inter-edge Consistency.}

      We now prove that  Modularity is not Connective.
      If it were, then by Lemma \ref{lemma:connectivity-no-cutedge}, there would be a value $n_0$ so that all clusters of size at least $n_0$ have min cut size greater than $1$, and so do not have any cut edge.
      Let $n$ be picked so that $2n \geq n_0$, and
       consider the network given in Appendix Lemma \ref{lemma:appendix-modularity-interedge} where $G_1$ is a component with $2n$ vertices containing two $n$-cliques connected by an edge and $G_0$ is a sufficiently large $p$-star, so that the optimal Modularity clustering returns $G_1$ as a single cluster. 
      Note that $G_1$  has a cut-edge, so that its minimum cut size is $1$. This contradicts our assumption, proving  that Modularity violates  connectivity.   

      \textcolor{black}{Finally, we prove that Modularity fails the Pair-of-Cliques component.  As shown in Corollary \ref{corollary:appendix-two-networks}, for any $k \geq 5$, a network that two components, with one a Pair-of-Cliques component where the cliques are of size $\binom{k}{2}$ and the other a clique of size $2\binom{k}{2}+1$, the optimal modularity clustering will return the two components as the two clusters. Thus, Modularity fails the Pair-of-Cliques axiom.}
\end{proof}

    \subsection*{Theory for IKC} 

Recall that we examine two versions of IKC: the ``default" setting that enforces a positive modularity score on all its non-singleton clusters, and the other, which we refer to as IKC(no-mod), that does not. 
Here we present the theory specifically for the default usage of IKC.

\begin{figure*}
\centering
\begin{subfigure}[b]{0.5\textwidth}
  \centering
   \includegraphics[width=0.9\textwidth]{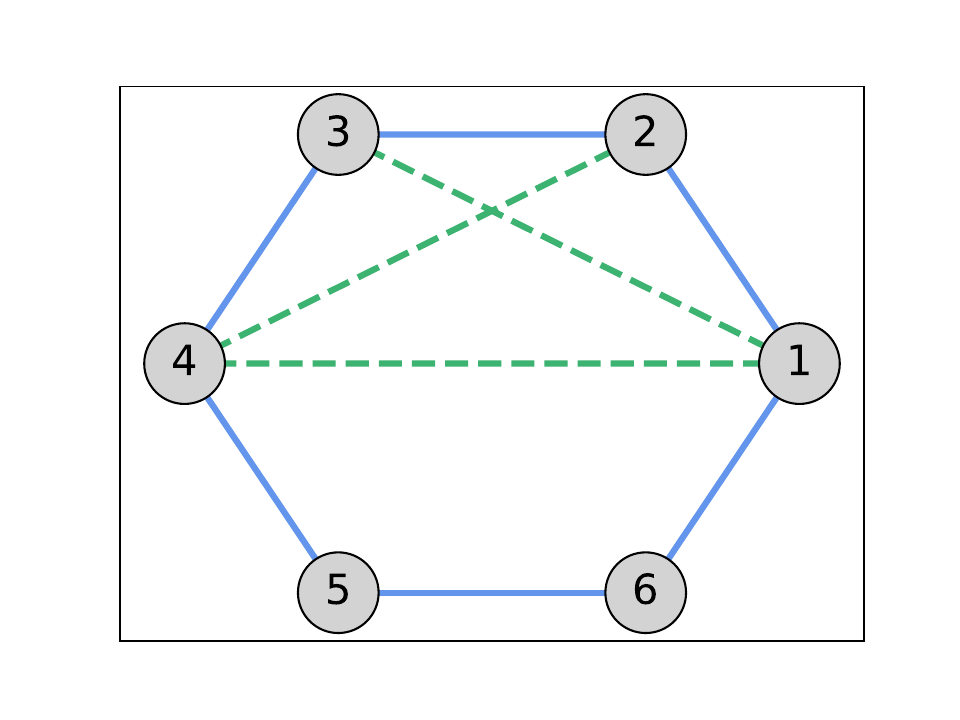}
  \caption{$N_1$, green edges are added to create $N_1'$.}
  \label{fig:IKC-standard-consistency}
\end{subfigure}%
\hfill
\begin{subfigure}[b]{0.5\textwidth}
  \centering
  \includegraphics[width=0.9\textwidth]{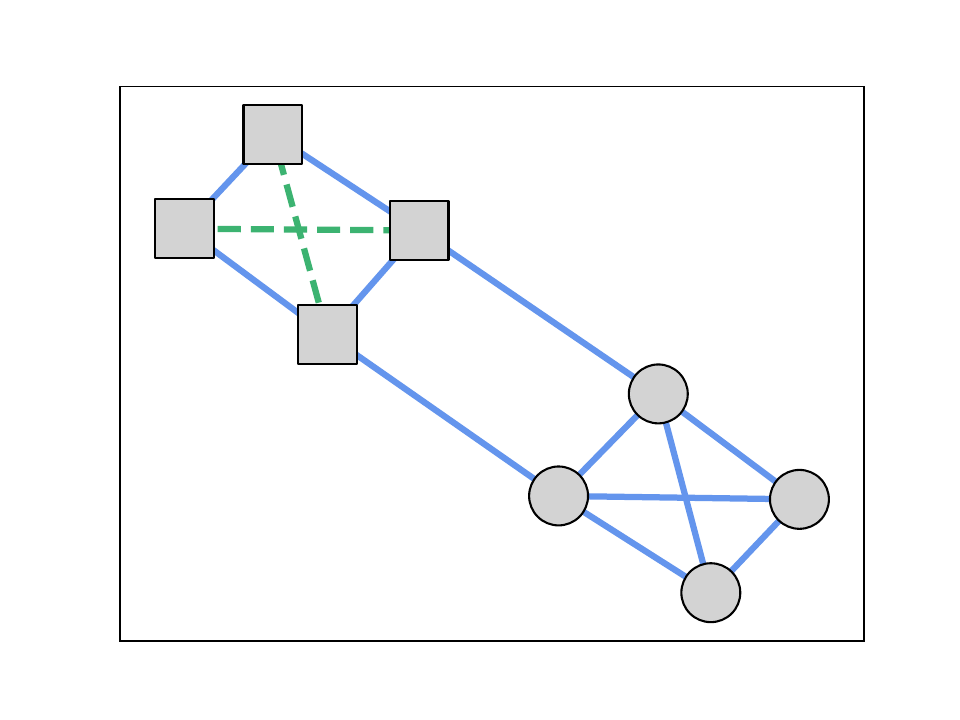}
  \caption{$N_2$, green edges are added to create $N_2'$}
  \label{fig:IKC-refinement-consistency}
\end{subfigure}
\hfill
\begin{subfigure}[b]{0.5\textwidth}
\centering
\includegraphics[width=0.9\textwidth]{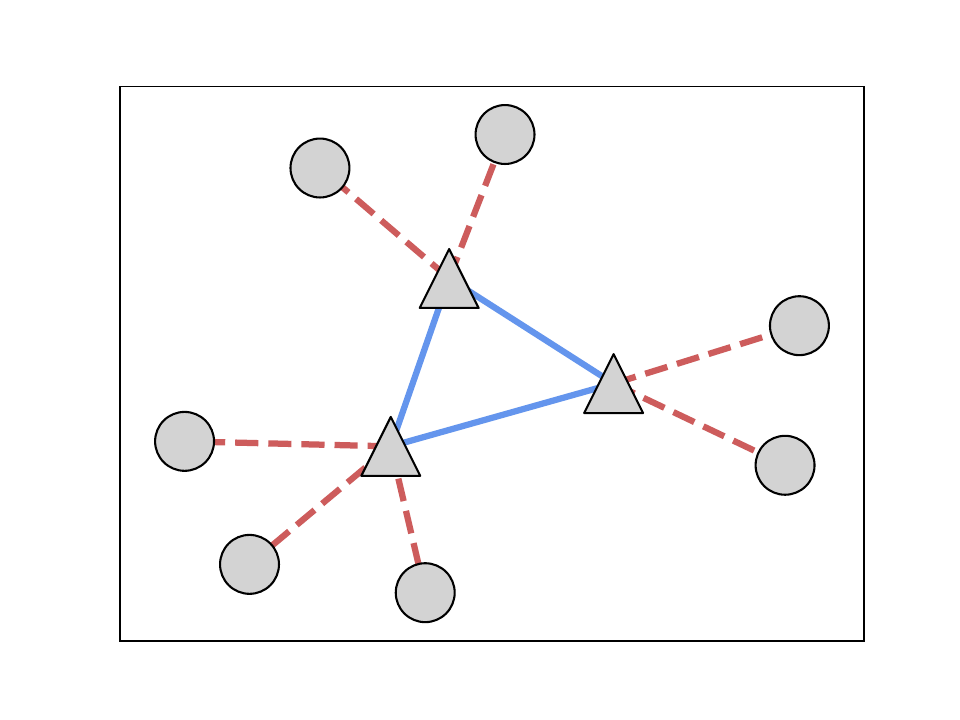}
\caption{$N_3$, red edges are deleted to create $N_3'$ }
\end{subfigure}
\caption{\textbf{Theoretical properties of IKC and IKC(no-mod).}
In each case, the network shown is one component of a network with two components, where the second component is a single edge (hence the shown component always has positive modularity).
Green edges are added to a starting network, red edges are deleted from a starting network, and blue edges represent edges that are in the starting and final network.
Subfigure (a) gives one component in a network $N_1$ where IKC and IKC(no-mod) both  fail Standard Consistency. 
Subfigure  (b) gives one component in a network $N_2$ where IKC and IKC(no-mod) both fail Refinement Consistency.  Subfigure (c) gives an example of one component in a network $N_3$ where IKC  will return only singleton clusters (due to its check for positive modularity). 
However, if the red edges were deleted, then IKC would return the 3-clique, establishing IKC fails Inter-Edge Consistency.}
\label{fig:IKC-fails-two}
\end{figure*}
  
\begin{theorem}
    IKC violates the Richness, Standard Consistency, Refinement Consistency, Inter-edge Consistency, Connectivity, Pair-of-Cliques,  \textcolor{black}{and Fixed Point axioms}.
    \label{theorem:ikc-axioms}
\end{theorem}
    
    \begin{proof}
    To see that Richness is violated, note that  a clustering containing every vertex in a network has a Modularity score of zero, and thus can never be considered a valid cluster for any edge set.    
    The proofs for IKC violating Standard Consistency, Refinement Consistency, and Inter-edge Consistency are based on networks shown in Figure \ref{fig:IKC-fails-two}.  In each subfigure, the shown graph is one component of a two-component network, where the other component is a single edge. The edge colors in each subfigure indicate edges that are present (blue), present but deleted (red), or not present but will be added (green).

    For the proof that IKC violates  Standard Consistency, we refer to Figure \ref{fig:IKC-fails-two}(a).
   This figure describes a  network $N_1$ with blue edges with two components, where one of these components is a simple 6-cycle and the other component is a single edge; the green edges are added to define a modified network $N_1'$. 
    Running IKC on $N_1$ would return the shown 6-cycle component as a cluster, since it is a $2$-core and has positive Modularity.  
    In $N_1'$, the vertex set $\{1,2,3,4\}$ forms a 3-core that has positive Modularity, and there is no 4-core in $N_1'$.
    Hence, when IKC is applied to $N_1'$, the 3-core
    $\{1,2,3,4\}$ would be returned as the  cluster found in the first iteration. Therefore, the IKC output clustering has been changed by the addition of edges within a cluster, and so IKC violates Standard Consistency.

    To see that IKC violates Refinement Consistency, see Figure 
    \ref{fig:IKC-fails-two}(b).
    \textcolor{black}{The initial network $N_2$ contains only the blue edges and the final network $N'_2$ also contains the green edges. 
    In $N_2$, the round vertices form a 3-core that has positive Modularity, and there is no 4-core in $N_2$; therefore, the round vertices would be returned as a cluster by IKC when applied to $N_2$. After removing the 3-core of round vertices, the square vertices form a 2-core. and since they have positive Modularity, they would be returned as the second cluster by IKC when clustering $N_2$. }
    However, the network $N_2'$ has the \textcolor{black}{green edges added}.
    In $N_2'$, the component shown constitutes a 3-core that has positive Modularity and so would be returned as a cluster by IKC when applied to $N'_2$. Thus, IKC fails Refinement Consistency.

    We now show that IKC violates Inter-Edge Consistency. The network $N_3$ described in 
    Figure 
    \ref{fig:IKC-fails-two}(c)
    has two components, a component with a single edge, which we will refer to as $C_1$, and the displayed 10-node component with both blue and red edges, which we will refer to as $C_2$. 
   In the first iteration of IKC applied to $N_3$, the blue edge 3-clique is detected as a 2-core, but since its modularity score is not positive (specifically, its modularity score is $3/11 - (13/22)^2 < 0$), it would not be returned as a cluster, and its three nodes would be turned into singleton clusters. 
    Hence, on network $N_3$, IKC \textcolor{black}{will return only one non-singleton cluster, and that is the two nodes in $C_1$, and will not return any non-singleton clusters for  component $C_2$.} 
    This establishes that the red edges shown in this figure go between different clusters obtained by IKC on $N_3$. 

   Recall that the network $N_3'$ is formed by deleting the red edges from $N_3$, and so has  the same vertex set, with one component a 3-clique, one component a single edge, and then seven isolated nodes.
   When IKC is applied to $N_3'$, it would find the $3$-clique, and since it has positive modularity (specifically, its modularity score is $3/7 - (3/7)^2$), it would return the $3$-clique as a cluster.
  In other words, we have shown that deleting edges between different clusters changed what is returned by IKC, which contradicts the Inter-Edge Consistency Axiom.

    We provide a proof by contradiction that IKC violates Connectivity. Suppose it did; then for some function $f$ that is increasing unboundedly and for all $n$ and all clusters of size $n$ returned by IKC, the edge cut for the cluster will be of size at least $f(n)$. 
    Since $f(x) \rightarrow \infty$ as $x \rightarrow \infty$, this means that for some $n_0 \geq 2$, no clusters of size at least $n_0$ returned by IKC have cut-edges.
    Now consider a network with two components, where one component $C$ has two $n_0$-cliques connected by an edge and the other component contains a single edge. 
    This component $C$ is a $n_0-1$-core and has positive modularity, and the network does not contain any $n_0$-core. 
    Hence, IKC would return the component $C$ as the first cluster, and then the single edge as the second cluster.
    However, $C$ has a cut edge, violating our assumption, and establishing that IKC fails Connectivity.

\textcolor{black}{
We next consider whether IKC satisfies the Fixed Point axiom. Note that IKC requires that a returned cluster have positive modularity.
Therefore, if $C$ is a $k$-clique returned by IKC it has positive modularity within its network.
However, when IKC is reapplied to the cluster $C$, it calculates the modularity score with respect to $C$ as the entire network. Thus, $C$ will now have zero modularity score, and so will not be returned by IKC.
Therefore, IKC fails the Fixed Point axiom.}

We now
    show that IKC fails the Pair-of-Cliques axiom. \textcolor{black}{Consider a network that has at least two components, where the first has $n$-cliques  $A$ and $B$ connected by a single edge, and there is at least one edge not in the first component.  
    IKC would return this first component as a cluster since it has positive modularity, and so would fail to return the cliques $A$ and $B$ as clusters. 
    The other case is where the network has only one component, which has the two $n$-cliques connected by edges.  Since the modularity score of an entire network with a single component is $0$, IKC will return only singletons. 
    Thus, in both cases, IKC will fail to return the cliques $A$ and $B$ and will return only singletons. 
    Since this outcome holds for all values of $n$, this establishes that IKC fails the Pair-of-Cliques axiom.
    }
       
    \end{proof}

    \subsection*{Theory for IKC(no-mod)}
\begin{theorem}
     IKC(no-mod) satisfies the Richness, Inter-Edge Consistency, \textcolor{black}{and Fixed Point} axioms, but violates the Standard Consistency,   Refinement Consistency,  Connectivity, and Pair-of-Cliques  axioms. 
    \label{theorem:ikc-nomod-axioms}
\end{theorem}
\begin{proof}
    To establish Richness, we consider the same network as used in the proof of richness for Modularity  (Theorem \ref{theorem:modularity}), where every component is a clique.
    It is easy to see that when running IKC(no-mod), each component of the network is returned as a cluster, since every non-singleton component has positive Modularity and is a $k$-core for some value of $k$. Notably, IKC(no-mod) can return all vertices in a single cluster, as the positive modularity restriction has been removed. 

    The proofs for IKC violating Standard Consistency, Refinement Consistency, and Connectivity  do not rely on checking for positive Modularity, and so apply to IKC(no-mod). 
    It is trivial to see that IKC(no-mod) returns the Pair-of-Cliques component as a cluster, since it is an $(n-1)$-core (where each clique has $n$ nodes).
    Hence, IKC(no-mod) fails the Pair-of-Cliques axiom.
    \textcolor{black}{Finally, it is easy to see that because IKC(no-mod) does not check for positive modularity, IKC(no-mod) satisfies the Fixed Point axiom.}

    We now establish that IKC(no-mod) satisfies inter-edge consistency. 
    Consider two clusters  $c_1$ and $c_2$ returned by IKC(no-mod), with at least one edge between them, and assume
     $c_1$ a $k$-core and $c_2$  a $k'$-core, with $k \leq k'$.  Note that $k \neq k'$, as otherwise the connected subgraph on $c_1 \cup c_2$ would be a $k$-core and would be returned.
    Removing an edge $e$ connecting these two clusters would only affect the degree of nodes in  these two clusters, so all other clusters would remain unaffected by any edge deletion.
    Furthermore, after removing $e$,  $c_1$ would still be a $k$-core and $c_2$ would still be a $k'$-core.   
    Therefore, in running IKC on the network obtained  by deleting edge $e$ between the clusters $c_1$ and $c_2$, these sets would still be considered for being clusters, and since modularity is not evaluated, $c_1$ and $c_2$ would still be returned as clusters by IKC(no-mod).
    Moreover, since no other cluster is affected, IKC(no-mod) would return the same clustering on the resultant graph. Hence, 
    IKC(no-mod) follows inter-edge consistency.
    \end{proof}

\section*{Discussion}
\subsection*{Relationship to other work}

\textcolor{black}{
The closest related paper is \cite{van2014axioms}, who addressed axiomatic properties of graph clustering methods based on optimization problems when the graph has non-negative edge weights. 
van Laarhoven and Marchiori \cite{van2014axioms} propose two new axioms, Locality and Continuity, and also study   Monotonicity,  Richness, Permutation Invariance, and Scale Invariance, which are axioms proposed in \cite{kleinberg2002impossibility,ben2008measures}.
Of these, Monotonicity and Continuity are only relevant when the edge weights can be arbitrary non-negative real numbers.
They study seven clustering methods, including Components-are-Clusters, CPM, Modularity,  and two  variants of Modularity (fixed scale and adaptive scale modularity).
They establish  that Modularity- and CPM-optimization satisfy Richness and Continuity, but CPM-optimization satisfies Locality  and Modularity does not, thus showing an advantage to CPM-optimization. 
However, Adaptive scale modularity satisfies all the axioms they present, while CPM-optimization and Components-are-Clusters each fail one axiom (with Components-are-Clusters failing Continuity and CPM-optimization failing Scale Invariance). 
} 

\subsection*{Summary of theoretical results}

\textcolor{black}{Our evaluation of graph clustering methods with respect to the different axioms we posed provides insight into differences between the clustering methods.
In particular, one noteworthy outcome of this study is that CPM($\gamma$), i.e., optimizing under the Constant Potts Model, satisfies all seven axioms we study, for all $0 < \gamma < 1$.}
Hence,  unlike Kleinberg's axioms (which were designed for distance-based clustering), there is {\em no impossibility theorem for clustering of simple unweighted graphs in the distanceless context} \textcolor{black}{ on our set of axioms}.

\textcolor{black}{On the other hand,
every other method we studied fails at least two of the axioms.
We also see that Modularity---one of the most well known clustering methods---fails every axiom other than Richness, IKC run in default mode fails every axiom, and IKC(no-mod) fails four axioms.
In contrast to these properties of existing clustering methods, each of the toy clustering methods we studied satisfies at least five of the seven axioms.  
Thus, these axioms reveal differences between clustering methods, and provide potentially helpful guidance to users of clustering methods.}

Our study also provides some insight into which axioms are very easy to meet, and which ones are more likely to distinguish between methods.  
For example, Table \ref{tab:results_summary} shows that Richness is in general extremely easy to achieve, with only Nodes-as-Clusters and IKC run in default mode failing to meet this criterion.
The axioms based on consistency (i.e., Standard Consistency and its two relaxations) distinguish between methods, with Modularity and IKC in its default setting failing, but CPM, IKC(no-mod), and the two ``toy" clustering methods succeeding.  
Given that even the toy clustering methods satisfy this axiom, failure to achieve consistency can be seen as a clear indication of a weakness for Modularity and IKC in its default setting.
The results for Connectivity on the other hand show that only CPM and  Nodes-as-Clusters satisfy the axiom, revealing a basic weakness for all the other methods.

The results for the two axioms related to the resolution limit show large differences between methods, and require specific discussion.
However, here we note that  failing the Pair-of-Cliques axiom means that the clustering method can produce arbitrarily large clusters that have cut edges. 
Thus, the fact that Modularity, IKC, and IKC(no-mod) all fail the pair-of-cliques axiom means they inherently can return arbitrarily large but very poorly connected clusters.

\subsection*{The Resolution Limit}

The resolution limit, first established by Fortunato and Barth\'elemy in \cite{fortunato2007resolution} for Modularity optimization, was described in terms of having an optimal clustering failing to find communities (i.e., sets of nodes that had high modularity scores) that were contained in larger sets of nodes.    The example that was given was a ring-of-cliques, i.e., a graph consisting of a set of $n$-cliques, each adjacent to two other cliques by single edges, so they formed a ring.  Fortunato and Barth\'elemy showed that as the number of cliques  increased, an optimal clustering using Modularity Optimization would return two or more of the cliques for a given cluster, rather than the single cliques.
That Modularity would fail to return the cliques as the communities was clearly interpreted as a strong limitation of the method.

There are two somewhat separable aspects of the Resolution Limit as described by Fortunato and Barth\'elemy in \cite{fortunato2007resolution}: one is that under some conditions, the output set of clusters will not contain any clusters below some size (which may depend on the method), and the other is that there can be obvious communities that ought to be returned by the method, that fail to be returned.

 Traag et al.~\cite{traag2011narrow} posed a property, which we refer to as the Fixed Point Axiom, to address the Resolution Limit.  
To satisfy this property, a clustering method will not change the output when applied to a single cluster or set of clusters it produces.
In \cite{traag2011narrow}, any method that satisfied this property was said to be ``resolution limit free".

Our study shows that the the Fixed Point Axiom was often satisfied by the clustering methods we examined, and even by the two toy methods  Components-are-Clusters and Nodes-are-Clusters. 
Thus, clustering methods that produce clusters that are too small (i.e.,  Nodes-are-Clusters) or too large (i.e.,  Components-are-Clusters) can both satisfy the Fixed Point Axiom,  indicating that this axiom does not address the first of the two aspects of the Resolution Limit we identified.
Furthermore, neither of the two toy methods is able to detect cliques as the true clusters when they are properly contained in components within the network; hence, the Fixed Point Axiom does not address the second aspect we identified. 
In other words, the Fixed Point Axiom does not adequately characterize methods that satisfy the two objectives of being resolution-limit-free, according to our interpretation of the findings in \cite{fortunato2007resolution}.

Given how the Fixed Point Axiom does not adequately address the Resolution Limit issues as identified in \cite{fortunato2007resolution}, we formulated a simple test, called the ``Pair-of-Cliques" axiom. We say that a method satisfies the Pair-of-Cliques axiom if any time the network contains a component that has two sufficiently large cliques connected by an edge (where the minimum size depends on the method),  it will  return the individual cliques. 
We found that of the clustering methods we examined,  only CPM-optimization satisfied the Pair-of-Cliques axiom.
Moreover, as shown in Table \ref{tab:results_summary}, four methods satisfy the Fixed Point Axiom but fail the Pair-of-Cliques axiom.
This shows that the two axioms -- Pair-of-Cliques and Fixed Point -- are very different from each other, although both aim to address the Resolution Limit.

\textcolor{black}{
Part of the focus of the study \cite{van2014axioms} by van Laarhoven and Marchiori is the resolution limit, and in particular the Fixed Point Axiom proposed by Traag et al. \cite{traag2011narrow}.
They propose a new axiom,  Locality, and discuss its relationship to the Fixed Point Axiom (showing it is both stronger in some ways and weaker in others). 
They define Adaptive scale modularity 
as a modification to Modularity and prove that it satisfies Locality. 
However, they prove that Locality is also satisfied by   Components-are-Clusters, and it is easy to see that it is satisfied by  Nodes-are-Clusters, each of which fails the  Pair-of-Cliques axiom.
Thus, like the Fixed Point axiom, their Locality axiom does not fully address the issues raised in \cite{fortunato2007resolution} about the Resolution Limit.
}

\section*{Conclusion}

Motivated by \cite{kleinberg2002impossibility}, which  established impossibility theorems for clustering when the input is an $n \times n$ distance matrix, we examined the question of axiomatic clustering when the input is a simple unweighted graph  without a corresponding distance matrix.  We  introduced seven axioms for distanceless graph partitioning, with four based on Kleinberg's axioms.
We established that unlike Kleinberg's axioms, there is no impossibility theorem for our axioms.  Moreover, we showed that optimizing under the constant Potts model (CPM), the default criterion for the Leiden software \cite{traag2019louvain},  one of the most popular methods for large-scale graph partitioning, has stronger theoretical guarantees than the other clustering methods we examined.

The results here are focused on theoretical properties of methods, but they also shed light on empirical performance.
For example, satisfying connectivity depends only on presenting {\em some} function $f$ so that all clusters of $n$ nodes have min cuts greater than size $f(n)$. 
In our proof that CPM($\gamma$) satisfies connectivity, the function $f$ we provided depended on $\gamma$, with the consequence that it provides a very weak bound when $\gamma$ is small.
The dependence on $\gamma$ is investigated in greater depth in Lemma \ref{lemma:appendix-cpm-return-component-small-gamma}, where we showed that for a given network $N$, $\gamma$ can be chosen small enough so that the clusters are the components of $N$.
This theoretical weakness is also reflected in empirical studies, as observed by \cite{park2023identifying-conf}, which demonstrated that using Leiden for CPM-optimization with
 very small values for $\gamma$ resulted in relatively sparse clusters that can be poorly connected, and can even be trees.  
\cite{traag2011narrow} also presents a discussion of this issue for its impact on CPM-optimal clustering.
Given that in practice small values for $\gamma$ are often used in order to achieve high node coverage, this a non-trivial issue (see discussion in \cite{park2023identifying-conf}).

Our study also revealed that the concerns raised in \cite{fortunato2007resolution} regarding the resolution limit are not fully addressed by the definition of ``resolution-limit-free'' given in \cite{traag2011narrow}. 
Our simple ``pair-of-cliques'' axiom is an initial step towards investigating the resolution limit for clustering methods, but only gives one simple case that should be checked.
A more complete analysis is needed, but this is challenging since at the heart of the resolution-limit is the concept that {\em some communities are clear}, so that recovering them must be achieved by a good clustering method.
Unfortunately, characterizing what constitutes an obvious community is difficult, since defining these based on (say) having a positive modularity score  is clearly insufficient.
Thus, this is another direction for future work.

We leave several questions for future research. 
Other graph partitioning methods beyond Modularity, CPM, and IKC, should be evaluated for their axiomatic properties, and 
variants of graph partitioning methods that enforce edge-connectivity, as studied in  \cite{park2023identifying-conf}, should also be considered. 

\textcolor{black}{
In addition, \cite{park2023identifying-conf} presented the {\em Connectivity Modifier}, an approach for modifying an existing output clustering to ensure that all clusters are well-connected, according to a user-specified lower-bound on the minimum edge cut size for a given cluster.
Such  a modification, paired with (say) CPM-optimization (in which $\gamma$ is not fixed in advance) might lead to new clustering methods that have strong theoretical properties. It is easy to see that this modification would ensure that the clustering algorithm satisfies the Connectivity axiom and would not change whether the method satisfied Richness, but it is less clear that the modified method would still satisfy Standard Consistency, Refinement Consistency, or Inter-edge Consistency. These questions merit investigation.}

\bibliography{ref}
\clearpage

  \section*{Appendix}
  This section contains  additional proofs.

  \subsection*{Additional proofs for Modularity}
   We will use the following notation.
(i) If $X$ is a subset of the nodes in a network $N$, then $Q_X$ denotes the modularity score of the cluster $X$ within a clustering.
(ii)  If $\mathcal{C}$ is a clustering of a network $N$, then 
    $Score(\mathcal{C})$ denotes the total modularity scores of the clusters in the clustering.
 (iii) The  largest modularity score across all clusterings of a network $N$ is written as $Modularity(N)$. 
(iv) If $G \subset N$ is a subgraph of network $N$, then the largest modularity score of $G$ across all clusterings of $N$ that make $G$ either into a single cluster or a collection of clusters is denoted by $Modularity_N(G)$.

\begin{lemma}
    Let $G_1$ be a component in network $N$, where 
    $G_1$ consists of two  node-disjoint cliques $A$ and $B$, each with $n > 5$ nodes, and a single edge connecting nodes in the two cliques.
    There are only two options for how $G_1$ is clustered in a modularity-optimal clustering of $N$: either $G_1$ is returned as a single cluster, or $G_1$ is split into two clusters, $A$ and $B$.
    \label{lemma:appendix-no-split-cluster}
\end{lemma}
\begin{proof}   
    To demonstrate that splitting the nodes in these two cliques apart is never optimal, we take a look at \cite{belyi2022network}. In Theorem 1, \cite{belyi2022network} proves that two endpoints of an edge will be in the same cluster if these endpoints are identically connected to every other node in the network. With this theorem, most of possible partitions splitting the cliques can be discarded; however, there is a single exception. Say edge $(a_0,b_0)$ is the edge connecting the two $n$-cliques $A$ and $B$; a partition separating $a_0$ from the clique $A$ is contained in might still be valid, when only considering this Theorem, as $a_0$ is connected to $b_0$, while this is not true for any other node in $A$. This leaves us with several cases we still need to consider.
    Let  $A_0 = A \setminus \{a_0\}$ and $B_0 = B \setminus \{b_0\}$. Then the options for clustering that we must consider are:
  Option 1: $A_0$, $\{a_0\}$, $B_0$, and $\{b_0\}$. Option 2:
        $A_0$, $\{a_0\}$, and $B$.
        Option 3:  $A$, $B_0$, and $\{b_0\}$.
        Option 4:  $A_0$, $B_0$, and $\{a_0, b_0\}$. 
        Option 5: $A \cup \{b_0\}$ and $B_0$.
        Option 6:  $A_0$ and $B \cup \{a_0\}$.\\

\noindent{\textbf{Ruling out Clustering options 1--3. }}
    First we show that $Q_A > Q_{A_0} + Q_{\{a_0\}}$; this will establish also that $Q_B > Q_{B_0} + Q_{\{b_0\}}$.
    Hence, we will be able to rule out clustering options 1--3.

    Let $E$ denote the total number of edges in the network.
    Then the modularity of clusters $A_0$ and $\{a_0\}$ (and hence also of $B_0$ and $\{b_0\}$) can be written as:
    \begin{align*}
        Q_{A_0} &= \frac{\binom{n-1}{2}}{|E|} - \frac{(n-1)^4}{4|E|^2}\\
        Q_{\{a_0\}} &= -\frac{(n + 1)^2}{4|E|^2}
        \intertext{We then write the modularity of $A$ (and $B$) as:}\\
        Q_{A} &= \frac{\binom{n-1}{2} + n - 1}{|E|} - \frac{\left[(n-1)^2 + n\right]^2}{4|E|^2}\\
        \intertext{and with some arithmetic we get:}\\
        Q_A &= \frac{\binom{n-1}{2} + n - 1}{|E|} - \frac{(n-1)^4 + 2n(n-1)^2 + n^2}{4|E|^2}\\
        &= \frac{\binom{n-1}{2} + n - 1}{|E|} - \frac{(n-1)^4}{4|E|^2} - \frac{2n(n-1)^2 + n^2}{4|E|^2}\\
        &= \frac{\binom{n-1}{2}}{|E|} - \frac{(n-1)^4}{4|E|^2} + \frac{n-1}{|E|} - \frac{2n(n-1)^2 + n^2}{4|E|^2}\\
        \intertext{which is equivalent to:}\\
        Q_A &=  Q_{A_0} + \frac{n-1}{|E|} - \frac{2n(n-1)^2 + n^2}{4|E|^2}\\
    \end{align*} 
    
        Thus to determine whether a clustering in which   $A$ appears as a cluster  has a better modularity score than the clustering obtained by  splitting $A$ into two clusters, $A_0$ and $\{a_0\}$, we evaluate the conditions under which $Q_A > Q_{A_0} + Q_{\{a_0\}}$.
        This is equivalent to showing  
        \begin{align*}
            Q_{A_0} + \frac{n-1}{|E|} - \frac{2n(n-1)^2 + n^2}{4|E|^2} > Q_{A_0} - \frac{(n + 1)^2}{4|E|^2}\\
            \intertext{which is equivalent to:}
            \frac{n-1}{|E|} - \frac{2n(n-1)^2 - (n + 1)^2  + n^2}{4|E|^2} > 0\\
            \intertext{which is equivalent to:}
            4|E|(n-1) > 2n(n-1)^2 - (n + 1)^2  + n^2\\
        \end{align*} 
        This is always true, as we now argue. Note that  $E$ is the set of edges in the network and so $|E| \geq n^2-n+1$ (the number of edges in component $G_1$)  and so $4|E|(n-1) \geq 2n(n-1)^2$. Note also that $n^2 - (n + 1)^2$ is always negative. Therefore $Q_A > Q_{A_0} + Q_{\{a_0\}}$ and $Q_B > Q_{B_0} + Q_{\{b_0\}}$.
        As a result, clustering options 1 -- 3 can be eliminated.\\

\noindent{\textbf{Ruling out clustering option 4. }} 
        The modularity of $\{a_0, b_0\}$ can be written as:

        \begin{align*}
            Q_{\{a_0, b_0\}} &= \frac{1}{|E|} - \frac{n^2}{|E|^2}\\
            \intertext{We see that $Q_A > Q_{A_0} + Q_{\{a_0, b_0\}}$ if and only if}\\
            \frac{n-1}{|E|} - \frac{2n(n-1)^2 + n^2}{4|E|^2} + Q_{A_0} &> \frac{1}{|E|} - \frac{n^2}{|E|^2} + Q_{A_0}\\            
             \intertext{if and only if}\\
             \frac{n-1}{|E|} - \frac{1}{|E|} &> \frac{2n(n-1)^2}{4|E|^2} - \frac{3n^2}{4|E|^2}\\
              \intertext{if and only if}\\
            4|E|(n-2) &> 2n(n-1)^2 -  3n^2\\
            \intertext{And since $-3(n-1)^2 > -3n^2$, it follows that $Q_A > Q_{A_0} + Q_{\{a_0, b_0\}}$ if}\\
            4|E|(n-2) &> (2n - 3)(n-1)^2\\
                      &= (2n - 3)(n-2)^2 + (2n - 3)^2\\
                      &= (2n - 3)(n-2)^2 + 
                      (4n^2 - 12n + 8) + 1\\
                      &= (2n - 3)(n-2)^2 + 
                      (4n - 4)(n - 2) + 1 \\
            \intertext{If both sides are divided by $(n - 2)$ we get}\\
            4|E| &> (2n - 3)(n - 2) + 4n - 4 + \frac{1}{n-2}\\
                 &= 2n^2 - n  +  \frac{1}{n-2} + 6\\
        \end{align*}

Note that this is always true. Hence, we have established $Q_A > Q_{A_0} + Q_{\{a_0,b_0\}}$.
Therefore, we can rule out option 4.\\

        \noindent{\textbf{Ruling out clustering options 5 and 6 }}

        To eliminate the final options, 5 and 6, we show that $Q_A + Q_B > Q_{A \cup \{b_0\}} + Q_{B_0}$. We write $Q_A$ as
        \begin{align*}
            Q_A &= \frac{\binom{n}{2}}{|E|} - \frac{\left[2\binom{n}{2} + 1 \right]^2}{4|E|^2}\\
            \intertext{and additionally,}\\
            Q_{A \cup \{b_0\}} &= \frac{\binom{n}{2} + 1}{|E|} - \frac{\left[2\binom{n}{2} + n + 1\right]^2}{4|E|^2}\\
            \intertext{Since $Q_B=Q_A$ and $Q_{B_0} = Q_{A_0}$, we know }\\
            Q_B - Q_{B_0} &= \frac{n-1}{|E|} - \frac{2n(n-1)^2 + n^2}{4|E|^2}\\           
        \end{align*}
        Hence, $Q_A + Q_B > Q_{A \cup \{b_0\}} + Q_{B_0}$  if and only if
        \begin{multline}
            \left(\frac{\binom{n}{2}}{|E|} - \frac{\left[2\binom{n}{2} 
            + 1 \right]^2}{4|E|^2}\right) - \left(\frac{\binom{n}{2} + 1}{|E|} - \frac{\left[2\binom{n}{2} + n + 1\right]^2}{4|E|^2}\right) \\ + 
            \left(\frac{n-1}{|E|} - \frac{2n(n-1)^2 + n^2}{4|E|^2}\right) > 0
       \end{multline}
        Simplifying,  $Q_A + Q_B > Q_{A \cup \{b_0\}} + Q_{B_0}$  if
        \begin{multline}
            0 < \frac{4\binom{n}{2}^2 + 4\binom{n}{2} + 1}{4|E|^2} + \frac{4\binom{n}{2}^2 + 4\binom{n}{2}(n + 1) +  (n + 1)^2}{4|E|^2} 
            \\ - \frac{1}{|E|} + \frac{n-1}{|E|} - \frac{2n(n-1)^2 + n^2}{4|E|^2}\\
            = \frac{4n\binom{n}{2} + (n + 1)^2 + 1}{4|E|^2} + \frac{n-2}{|E|} - \frac{2n(n-1)^2 + n^2}{4|E|^2}\\
            = \frac{3n^2 + 2}{4|E|^2} + \frac{n - 2}{|E|} - \frac{2n^3 + 5n^2 - 2n}{4|E|^2}\\
            = \frac{n - 2}{|E|} - \frac{2n^3 + 2n^2 - 2n - 2}{4|E|^2}\\
      \end{multline}
        if and only if
        \begin{align*}
            4|E|(n - 2) &> 2n^3 + 2n^2 - 2n - 2\\
            &= (n - 2)(2n^2 + 6n + 10) + 18
            \intertext{Dividing both sides by $(n-2)$ gives us that $Q_A + Q_B > Q_{A \cup \{b_0\}} + Q_{B_0}$  if}\\
            4|E| &> 2n^2 + 6n + 10 + \frac{18}{n - 2}\\
            \intertext{Since $|E| \geq n^2 - n + 1$, we see that $Q_A + Q_B > Q_{A \cup \{b_0\}} + Q_{B_0}$  if }\\
            4n^2 - 4n + 4 &> 2n^2 + 6n + 10 + \frac{18}{n - 2}\\
            \intertext{which is equivalent to}
            2n^2 - 10n - 6 - \frac{18}{n-2} &> 0\\
        \end{align*}
        This is true for $n > 5$. Thus we eliminate the final options, 5 and 6. 
        
        Therefore, for any network with this structure, optimizing modularity does not separate the nodes within the cliques   $A$ and $B$. The lemma follows.
\end{proof}

\begin{lemma}
    Modularity violates Standard and Refinement Consistency.
    \label{lemma:appendix-modularity-standard-consistency}
\end{lemma}

\begin{proof}
    Consider a  network $G = (V,E)$ with two components, $G_0$ and $G_1$,  with  $G_1$   as in  Lemma \ref{lemma:appendix-no-split-cluster}; thus,  $G_1$ contains two cliques $A$ and $B$, each with $e$ edges, connected by a single edge. Let  $E_{other}$ denote the edge set for the other component $G_0$.
    By Lemma \ref{lemma:appendix-no-split-cluster}, in a modularity-optimal clustering of this network, there are only two options for how $G_1$ is clustered: either as a single cluster (containing all the nodes in $G_1$) or as two clusters, $A$ and $B$.

    We define $Q_2$ to be the modularity score of $G_1$ when the clustering produces two clusters (i.e., each clique is considered a single cluster) and  $Q_1$ is the modularity score of $G_1$ when the entire component is considered a single cluster (thus, the index indicates how many clusters $G_1$ is split into).
    Equivalently, $Q_2 = Q_A + Q_B$ and $Q_1 = Q_{G_1}$.
    We are interested in understanding when $Q_1 > Q_2$, so that returning a single cluster for $G_1$ is preferable to returning $A$ and $B$ as single clusters.
    We find $\Delta Q = Q_1 - Q_2$, by referring to Equation 14 from \cite{fortunato2007resolution}\footnote{Using the notation from \cite{fortunato2007resolution}, 
    in our network,  $l_1 = l_2 = e$, $b_1 = b_2 = 0$ 
    and $a_1 = a_2 = \frac{1}{e}$ (since a single edge connects the two cliques)}. 
    Hence we obtain:
    \begin{equation*}
        \Delta Q = \frac{2|E| - 4e^2 - 4e - 1}{2|E|^2}
    \end{equation*}

    Note that $Q_1 > Q_2$ if and only if:
    \begin{equation}
        |E| > 2e^2 + 2e + \frac{1}{2}
    \end{equation} 

    This inequality can be rewritten as (by subtracting $2e + 1$):
    \begin{equation}
        |E_{other}| > 2e^2 - \frac{1}{2}
    \end{equation}
    Thus, the modularity score of the clustering where $G_1$ is one component is larger than the modularity score of the clustering where $G_1$ is two clusters  if and only if (6) holds.
    
    Now consider $Modularity(N)$, the score of the best achievable modularity clustering of $N$.  
    We write this as $Modularity(N) = Modularity(G_1) + Modularity(G_0)$,
    as we require that output clusters be connected.
    Recall  $Modularity(G_1) = \max(Q_1, Q_2)$ is ensured by Lemma \ref{lemma:appendix-no-split-cluster}.
     Hence $Modularity(N) = \max(Q_1, Q_2) + Modularity(G_0)$.

    Next we consider the component $G_0$. 
    We will let $G_0$ be a $p$-star   (i.e., a graph with a center node adjacent to $p$ other nodes that all have degree $1$).  
    Consider an optimal modularity clustering of $G_0$ within this network. If this clustering breaks $G_0$ into two or more clusters, then exactly one cluster contains the center node and all the other clusters are singletons (since we require that the clusters be connected).  
    Let $x$ be the number of singleton clusters (that do not include the center node), and assume the total number of nodes is $p+1$ (so there are $p$ nodes adjacent to the center node). Then the modularity score of this clustering is given by:
    \begin{align*}
        Q_{p-star} &= \frac{p - x}{|E|} - \left(\frac{2p - x}{2|E|}\right)^2 - x\left(\frac{1}{2|E|}\right)^2 \\
                &= \frac{p}{|E|} - \frac{x}{|E|} - \frac{4p^2 - 4px + x^2}{4|E|^2} - \frac{x}{4|E|^2} \\
                &=  \frac{4px - 4|E|x - x^2 - x - 4p^2 + 4|E|p}{4|E|^2} \\
    \end{align*}

    Note that this equation is maximized at $x = 0$, since $x \geq 0$ and $|E| \geq p$, so clustering the entire star into a single cluster has the optimal modularity score.

   We set up $G$ so that $G_0$ is a $p$-star in our original network (so that $G_0$ is returned as a cluster) and then we add edges until $G_0$ is a clique, creating a new network. 
  We can select values for $e$ (the number of edges in the cliques in $G_1$) and $p$ (where $G_0$ is a $p$-star)  that will cause
 Inequality (6) to be violated (and so indicate $Q_2 > Q_1$) in the case where $G_0$ is a $p$-star and not violated (and so indicate $Q_1> Q_2$) in the case where $G_0$ is a  $(p+1)$-clique.  
 This will prove that Modularity violates refinement consistency, and so also violates standard consistency.

    For instance, if $p = 2e$ (and recalling that $e \geq \binom{5}{2}$),
    then when $G_0$ is a $p$-star:
    \begin{equation*}
        |E_{other}| = 2e < 2e^2 - \frac{1}{2},
    \end{equation*}
    which violates Inequality (6), and hence means that $G_1$ will be split into two clusters, $A$ and $B$, in an optimal modularity clustering.
    However, when $G_0$ is a clique: 
    \begin{equation*}
        |E_{other}| = \binom{2e + 1}{2} = \frac{2e(2e+1)}{2} = 2e^2 + e > 2e^2 - \frac{1}{2}
    \end{equation*}
    which obeys Inequality (6).
    Note that this argument applied to all  $e \geq \binom{5}{2}$.
    
    To summarize,  we see that returning $A$ and $B$ as separate clusters is modularity-optimal in the case where $G_0$ is a $p$-star,
    whereas returning $G_1$ as a single cluster is modularity-optimal when $G_0$ is a $(p+1)$-clique.  
    This means that Modularity violates Standard Consistency.
\end{proof}

\textcolor{black}{The proof of Lemma \ref{lemma:appendix-modularity-interedge} yields this helpful lemma as an immediate corollary:}
 \begin{corollary}
 \textcolor{black}{
Let $e = \binom{k}{2}$ where $k \geq 5$ is a positive integer and let $N_1$ be a network with two components, $C_1$ and $C_2$, where $C_1$ is a Pair-of-Cliques component with two cliques $A$ and $B$ connected by an edge, with $A$ and $B$ each having exactly $e$ edges.
Let the other component $C_2$ of $N_1$ be a $p$-star, with $p = 2e$. 
Let $N_2$ also be a network with two components, $C_1$ and $C_3$ (i.e.,  $C_1$ is the same Pair-of-Cliques component as for $N_1$) and where $C_3$ is a $(p+1)$-clique.
Then the optimal modularity clustering of $N_1$ will return $A$ and $B$ as separate clusters, and an optimal modularity clustering of $N_2$ will return $C_1$ as a cluster.} 
\label{corollary:appendix-two-networks}
 \end{corollary}

\begin{lemma}
    Modularity fails the inter-edge consistency axiom.
    \label{lemma:appendix-modularity-interedge}
\end{lemma}

\begin{proof}

We form a network $N$ where  $G_1$ (a pair-of-cliques component) is one component,  and then we add a network $N'$ that is not connected to $G_1$, and that has the following properties: 
 \begin{itemize}
 \itemsep 0pt
 \item Property (1):
The optimal modularity clustering of $N = G_1 \cup N'$ returns $G_1$ as a single cluster, and splits $N'$ into at least one more cluster than the number of components in $N'$. 
\item Property (2):
$N'$ is minimal subject to Property (1), which means that if we delete {\em any} edge of $N'$, then $N$ longer satisfies Property (1). 
\end{itemize}
Now suppose such a network $N'$ exists (and note that $N'$  depends on the value of $n$, where $G_1$ has two $n$-cliques).   
Since $N'$ satisfies Property (2), if we delete any edge in $N'$ at all, then Property (1) does not hold.
Let  $\mathcal{C}$ be an optimal modularity clustering of $N$. 
Now consider the network $N^*$ produced by the deletion of an edge $e_0$ that goes between two different clusters in $\mathcal{C}$ (such an edge exists since the optimal clustering produces more clusters than there are components), and then running modularity on $N^*$ to produce clustering $\mathcal{C}'$. 
Since $N'$ was minimal subject to Property (1), it follows that Property  (1) does not hold for $N' \setminus \{e_0\}$ (the network produced by deleting the edge $e_0$ but not its endpoints from $N'$).
Hence, in the clustering $\mathcal{C}'$, either $G_1$ is not returned as a cluster or $N^*$ does not splits into at least two clusters. Therefore, no matter how $\mathcal{C'}$ differs from $\mathcal{C}$, it follows that Modularity violates inter-edge consistency.

Therefore, all we need to do to complete the proof is to establish that such a network $N'$ exists that satisfies Properties (1) and (2), above. 
Consider a network $N$ with two components. $G_1$ is made of two cliques of equal size (containing $e$ edges), connected by a single edge.  
We let $N'=G_0$, which  is made of two sets of vertices. The first set contains $e^2$ edges and the second contains $e^2 - 1$ edges; these two sets are connected by a single edge $e^*$; therefore, $|E_{G_0}| = 2e^2$.  
Specifically, we need to show  Property (1), i.e.,  that the optimal modularity clustering of $N = G_1 \cup G_0$ returns $G_1$ as a single cluster and splits $G_0$ into at least two clusters,  and Property (2), i.e.,  that the removal of any edge in $G_0$ will not satisfy Property  (1).
This will complete the proof.

Given the fact that $G_1$ and $G_0$ are components and we require that the clusters be connected, the modularity score for the entire network $N$ satisfies $Modularity(N) = Modularity_N(G_1) + Modularity_N(G_0)$.

By the proof of Lemma \ref{lemma:appendix-modularity-standard-consistency},
     $G_1$ will be clustered as a single cluster if and only if  $|E(G_0)| > 2e^2 - \frac{1}{2}$. Given how we have defined $G_0$ and $G_1$, this is equivalent to saying that $G_1$ will be returned as a single cluster if and only if $2e^2 > 2e^2 - \frac{1}{2}$. Hence for how we have defined the network, $G_1$ is returned as a single cluster in any optimal modularity clustering.

     We now show that if we do not remove edge $e^*$, then $G_0$ is clustered into at least two clusters in any modularity-optimal clustering of the network $N$, which will establish Property (1).
     We refer to Equation 14 from \cite{fortunato2007resolution}, where we define $\Delta Q = Q_4 - Q_3$ and $Q_3$ as the modularity score of the case where $G_0$ is split into two clusters across the single cut edge $e^*$ and $Q_4$ is the score for the case where $G_0$ is considered a single cluster.
     Note that if $\Delta Q < 0$ then returning a single cluster for $G_0$ is not modularity-optimal. 
      
    \begin{align*}
        \Delta Q &= \frac{2|E| - \left(2 + \frac{1}{e^2}\right)\left(\frac{1}{e^2 - 1} + 2\right)e^2\left(e^2 - 1\right)}{2|E|^2}\\
        &= \frac{2|E| - \left(\frac{2}{e^2-1} + 4 + \frac{1}{e^4 - e^2} + \frac{2}{e^2}\right)\left(e^4 - e^2\right)}{2|E|^2}\\
        &= \frac{2|E| - \left(\frac{4e^8 - 8e^6 + 3e^4 + e^2}{e^4 - e^2}\right)}{2|E|^2}\\
        &= \frac{2|E| - 4e^4 + 4e^2 + 1}{2|E|^2}\\
    \end{align*} 
    Hence, 
    \begin{equation*}
     \Delta Q < 0 \textbf{ if and only if }   |E| < 2e^4 - 2e^2 - \frac{1}{2}
    \end{equation*}
     
    We know that $|E| = 2e^2 + 2e + 1$.
    Hence, for all $e \geq 2, \Delta Q < 0$.
    Therefore, for the network $N$ we have constructed, the modularity-optimal clustering of $G_0$ has at least two clusters, and  we have established that $N'=G_0$ satisfies Property (1). 

    We now establish  that $N'=G_0$ satisfies (2).  Imagine our network $N^*$, where the  edge $e^*$ contained in $G_0$ is removed. According to the proof of Lemma \ref{lemma:appendix-modularity-standard-consistency}, $G_1$ will be returned as a single cluster if and only if $|E(G_0)| - 1 > 2e^2 - \frac{1}{2}$, which is the same as $2e^2 -1 > 2e^2 - \frac{1}{2}$, which is never true. Hence if we remove the edge $e^*$ in $G_0$, then $G_1$ will not be returned as a single cluster.

    Hence, $N'=G_0$  satisfies Properties (1) and (2) above, and the lemma is proven.  
\end{proof}

\subsection*{Additional theory for CPM}
\textcolor{black}{
The following lemma is not directly relevant to understanding the properties of CPM-optimization with respect to the axioms we stated, but sheds some light on the behaviour of CPM($\gamma$) and how this is impacted by $\gamma$.}
\begin{lemma}
    If $N$ is a network and $C$ is a component in the network,
    then for all sufficiently small $\gamma$, every optimal CPM($\gamma$) clustering returns $C$ as a cluster.
    Specifically, if $\gamma \binom{n}{2} < 1$ and $C$ is a component of size $n$ in a network $N$, then
$C$ will be returned as a cluster in every CPM($\gamma$)-optimal clustering.
    \label{lemma:appendix-cpm-return-component-small-gamma}
\end{lemma}
\begin{proof}
    
We begin by calculating the CPM($\gamma$) score of the cluster $C$. Letting $E(C)$ denote the edge set of $C$ and $n$ denote the number of nodes in $C$, we obtain:
\begin{equation*}
    CPM(C) = |E(C)|- \gamma \binom{n}{2}
\end{equation*}
Since the CPM function is continuous in $\gamma$,  as $\gamma \rightarrow 0$, this will become arbitrarily close to $|E(C)|$ (but is always  smaller).
Hence in particular, we can pick $\gamma$ small enough to produce
\begin{equation*}
    CPM(C) \geq |E(C)| - 1
\end{equation*}
Specifically, if $\gamma \binom{n}{2} <1$, the above equation holds. 

Let $\gamma_0$ be such a value,  and consider a clustering of $N$ that is optimal under CPM($\gamma_0$). 
Suppose that the optimal clustering of $N$  splits $C$ into $k \geq 2$ clusters, $C_1, C_2, \ldots, C_k$.  
Since $C$ is connected, there is at least one edge in $E(C)$ that is not in any cluster. 
Letting $m_i$ denote the number of edges in cluster $C_i$, the
 CPM score of this optimal clustering (for $C$) is given by 
\begin{equation*}
    \sum_{i=1}^k CPM(C_i) < \sum_{i=1}^k m_i \leq |E(C)|-1
\end{equation*} 
Note that the first inequality follows since $\gamma_0 >0$ is required, and the second inequality follows since at least one edge is not in any cluster. 
However,  this is strictly less than the CPM score of the cluster containing the entire component $C$, contradicting its optimality.
Hence, for small enough $\gamma$, the optimal CPM($\gamma$) clustering returns the entire component as a cluster.
\end{proof}

While CPM($\gamma$) is provably connective, the function $f$ that provides the guarantee depends on $\gamma$.
Now, suppose we ask instead: Is there a function $f: \mathbb{R}^+ \rightarrow \mathbb{R}^+$ that works for all $\gamma$, i.e., so that for all $\gamma$, the mincut size for every CPM-optimal cluster of size $n$ is greater than $f(n)$?
The answer is unfortunately {\em no}, as we now argue.

Suppose such a function $f$ were to exist.
In this case, we could pick a value for $n$ so that $f(n) =2$.
For that value of $n$, we would then pick $\gamma$ small enough so that $\gamma  \binom{n}{2} < 1$, with the consequence that every component of size $n$ would be returned as a cluster (Lemma \ref{lemma:appendix-cpm-return-component-small-gamma}). Since a component can contain a cut edge, this would contradict the assumption that $f(n) = 2$, so that the min cut size is at least $2$.

The consequence of this observation is that the connectivity guarantee provided for CPM($\gamma$) depends on $\gamma$, and that small values for $\gamma$ allow for large clusters with cut edges being returned.

\end{document}